\documentclass[english,aps,prl,notitlepage,twocolumn]{revtex4-1}
\usepackage[T1]{fontenc}
\usepackage{color}
\usepackage{amsmath}
\usepackage{amssymb}
\usepackage{graphicx}
\usepackage{bm}

\usepackage{footnote}

\usepackage{babel}

\usepackage{lipsum}

\makeatletter

\usepackage{amsmath}
\usepackage{endnotes}
\makeatother

\usepackage{babel}
\begin{document}


\title{Velocity Scanning Tomography for Room-Temperature Quantum Simulation}


\author{Jiefei Wang,$^{1,2,*}$ Ruosong Mao,$^{1,*}$ Xingqi Xu,$^{1,*,\dagger}$ Yunzhou Lu,$^{1}$ Jianhao Dai,$^{1}$ Xiao Liu,$^{1}$ Gang-Qin Liu,$^{3}$ Dawei Lu,$^{4}$ Huizhu Hu,$^{2}$ Shi-Yao Zhu,$^{1,2,5}$ Han Cai,$^{2,\dagger}$  and Da-Wei Wang$^{1,2,5,\dagger}$ \\}

\affiliation{$^1$Zhejiang Province Key Laboratory of Quantum Technology and Device, School of Physics, and State Key Laboratory for Extreme Photonics and Instrumentation, Zhejiang University, Hangzhou 310027, China\\
$^2$College of Optical Science and Engineering, Zhejiang University, Hangzhou 310027, China\\
$^3$Beijing National Laboratory for Condensed Matter Physics and Institute of Physics, Chinese Academy of Sciences, Beijing 100190, China\\
$^4$Shenzhen Institute for Quantum Science and Engineering and Department of Physics, Southern University of Science and Technology, Shenzhen 518055, China\\
$^5$Hefei National Laboratory, Hefei 230088, China
}
\date{\today}

\begin{abstract}
Quantum simulation offers an analog approach for exploring exotic quantum phenomena using controllable platforms, typically necessitating ultracold temperatures to maintain the quantum coherence. Superradiance lattices (SLs) have been harnessed to simulate coherent topological physics at room temperature, but the thermal motion of atoms remains a notable challenge in accurately measuring the physical quantities. To overcome this obstacle, we invent and validate a velocity scanning tomography technique to discern the responses of atoms with different velocities, allowing cold-atom spectroscopic resolution within room-temperature SLs. By comparing absorption spectra with and without atoms moving at specific velocities, we can derive the Wannier-Stark ladders of the SL across various effective static electric fields, their strengths being proportional to the atomic velocities. We extract the Zak phase of the SL by monitoring the ladder frequency shift as a function of the atomic velocity, effectively demonstrating the topological winding of the energy bands. Our research signifies the feasibility of room-temperature quantum simulation and facilitates their applications in quantum information processing.
\end{abstract}

\maketitle

Quantum platforms such as cold atoms, superconducting circuits, and ion traps are extensively employed for simulating exotic quantum matter thanks to their exceptional controllability and flexibility in engineering Hamiltonians \cite{Cooper2019,Ozawa2019,Monroe2021}. In simulating  emergent quantum phenomena \cite{Manin1980,Feynman1982,Georgescu2014}, one common requirement is the cooling of quantum systems to sufficiently low temperatures to eliminate the thermal noise, which can otherwise disrupt the coherence of quantum states.
However, for practical applications of quantum simulation, such as topological insulator lasers \cite{Harari2018,Bandres2018}, it becomes imperative for quantum platforms to operate at ambient temperatures. Unfortunately, apart from photonic systems, achieving such temperature resilience is quite uncommon. This challenge can be attributed to the energy scale of the quantum systems involved.
Photons typically possess energy quanta around 1 eV, considerably larger than the energy associated with blackbody radiation at room temperature. In contrast, cold atoms  used for quantum simulation rely on motional degrees of freedom with extremely small energy scales \cite{Greiner2002,Jotzu2014}, discernible only at ultracold temperatures below 1 $\mu$K. This limitation restricts the accessibility and potential applications of topological effects in atom-based systems.

To harness the strengths of atoms, characterized by flexible Hamiltonian engineering, along with the resilience to thermal noise exhibited by photons, superradiance lattices (SLs) \cite{Wang2015,Wang2015optica} based on the collectively excited states of atoms \cite{Scully2006,Scully2009,He2020,Araujo2016,Roof2016} have been used to conduct quantum simulations at room temperature \cite{Cai2019,He2021,Xu2022,Mao2022}. In these experiments, multiple laser beams are employed to couple the internal electronic energy levels of an ensemble of atoms, forming momentum-space tight-binding lattices, where the atomic motion introduces an effective static electric field, with its strength proportional to the atomic velocity. However, one significant challenge remains. The thermal distribution of atomic velocities presents an obstacle to obtaining the dynamic response of the SL in different effective electric fields. This limitation hinders the precise extraction of geometric phases and quantum simulation of higher-dimensional topological matter at room temperature.


\begin{figure*}[htp!]
    \centering
    \includegraphics[width =1\linewidth]{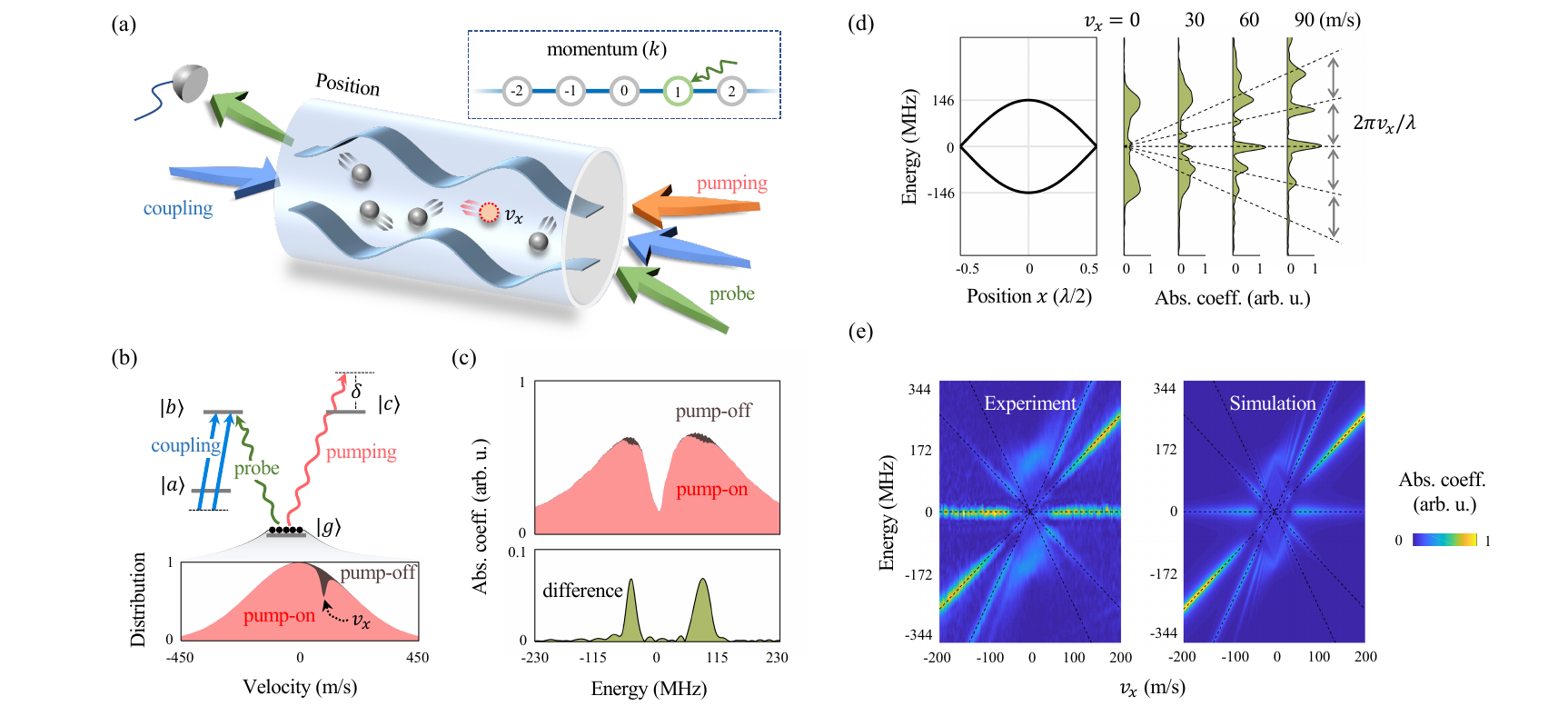}
    \caption{\textbf{Velocity scanning tomography.} (a) Schematics of the experimental set-up. Two  counter-propagating lasers (blue) couple atoms to form an SL (inset, the number on each site labels the momentum of the corresponding TDS, in the unit of $k$), whose DOS is detected by a probe field (green). A pumping laser (red) is used to burn a hole in the thermal atomic velocity distribution. (b) The schematic coupling between the atomic states and lasers (up) and a typical hole-burnt atomic velocity distribution (pink) compared to the Maxwell distribution of thermal atoms (gray). The central velocity of the pumped atoms is $v_x=\delta\lambda/2\pi$. (c) Experimentally measured absorption spectra without (gray) and with (pink) the pumping laser. Their difference characterizes the contribution from the pumped atoms. (d) Left panel: the dispersion relation of a gapless two-band SL. Right panel: the measured absorption spectra for the SL of atoms with different $v_x$. (e) The velocity-resolving energy spectra of SLs, characterized by the WSLs (discrete peaks with equal energy spacing $2\pi v_x/\lambda$). The parameters $t_1=t_2=73$ MHz and $\Delta=0$ in (d) and (e). {Dashed lines are the theoretically predicted positions of the WSLs}. 
    }
    \label{fig1}
\end{figure*}



In order to effectively discern responses from atoms with different velocities, we have invented a technique termed as velocity scanning tomography (VST). This method enables the measurement of SL absorption spectra for atoms moving at a selected velocity. A key ingredient of VST is to pump atoms that move with a selected velocity to another energy level, achieved using a laser with a narrow linewidth. This spectral hole-burning technique breaks up the room-temperature SLs into a set of cold-atom SLs \cite{Chen2018, Mi2021, Wang2020npj} in different effective electric fields. It can also be used to resolve inhomogeneously broadened quantum emitters such as those in solid-state experiments \cite{Riedmatten2008, Afzelius2010, Putz2017,Lei2023}. The SL response of the pumped atoms can be obtained by contrasting the outcomes with and without such pumping. By systematically scanning through these velocities, we are able to monitor the energy shift of the Wannier-Stark ladders (WSLs) \cite{Wannier1960,Wilkinson1996,Sundaram1999,Lee2015} of the SL, corresponding to changing the strength of an effective electric field.
The VST procedure empowers us to acquire velocity-dependent spectra of thermal atoms, where the precision in distinguishing atomic velocities is primarily constrained by the inherent linewidth of the atomic transition. From these spectra, we can accurately derive the Zak phase \cite{Zak1989,Xiao2010} of the SL under varying lattice parameters. Consequently, we can obtain Chern numbers by observing the topological winding of the Zak phase through dimension reduction techniques \cite{Qi2008,Yu2011,Soluyanov2011,Taherinejad2014,Alexandradinata2014}.

The experiment is conducted in a natural abundance rubidium vapor cell (Fig.~\ref{fig1}(a)), which is 2-centimetre long and shielded from external magnetic fields. We use a weak probe laser and two strong counter-propagating lasers (along the $x$-axis) to couple atoms in a $\Lambda$-type electromagnetically induced transparency (EIT) configuration (see Fig.~\ref{fig1}(b)). The probe laser couples the transition between the ground state $|g\rangle \equiv |5^2S_{1/2},F=1\rangle$ and the excited state $|a\rangle \equiv |5^2P_{1/2},F=2\rangle$ of the $^{87}$Rb atoms, while the coupling lasers couple the excited state $|a\rangle$ to a metastable state $|b\rangle \equiv |5^2S_{1/2},F=2\rangle$, with Rabi frequencies $t_1$ and $t_2$ and detuning $\Delta$. Such couplings between atoms and lasers construct a tight-binding SL of timed Dicke states (TDS),  as shown in the inset of Fig.~\ref{fig1}(a). The SL contains two sublattices, with $t_1$ and $t_2$ being the inter- and intra-sublattice coupling strengths and $\Delta$ being the energy off-set between the two sublattices.

To isolate atoms with varying velocities, we employ a narrow linewidth bleaching laser to excite atoms with a specific velocity from the ground state $|g\rangle$ to an auxiliary excited state $|c\rangle \equiv |5^2P_{1/2},F=1\rangle$. The laser frequency, detuned by $\delta$ from the atomic transition, is adjusted to target atoms moving with a particular velocity. This laser selectively bleaches a subset of atoms, generating a narrow dip in the atomic velocity distribution at a central velocity $v_x = \delta\lambda/2\pi$, with $\lambda$ being the laser wavelength, as illustrated in Fig.~\ref{fig1}(b). The Lorentzian width of this dip is limited by the decay rate $\Gamma$ of the $|c\rangle$ state, approximately $2\Gamma\lambda/2\pi$, which is around 10 m/s for $^{87}$Rb atoms. This value is significantly smaller than the full width at half maximum (FWHM) of the thermal velocity distribution, which is approximately 350 m/s.
To distinguish the minute difference in the absorption spectra with and without the presence of the pumping laser (as shown in Fig. \ref{fig1}(c)), we utilize a shutter operating at a rate of 3 kHz to alternately enable and disable the bleaching laser. This periodic switching generates a phase-locked Fourier component that corresponds to the contribution of the atoms within the targeted velocity group (see Supplemental Material \cite{SM}).


\begin{figure}[htp!]
    \centering
    \includegraphics[width = 1\linewidth]{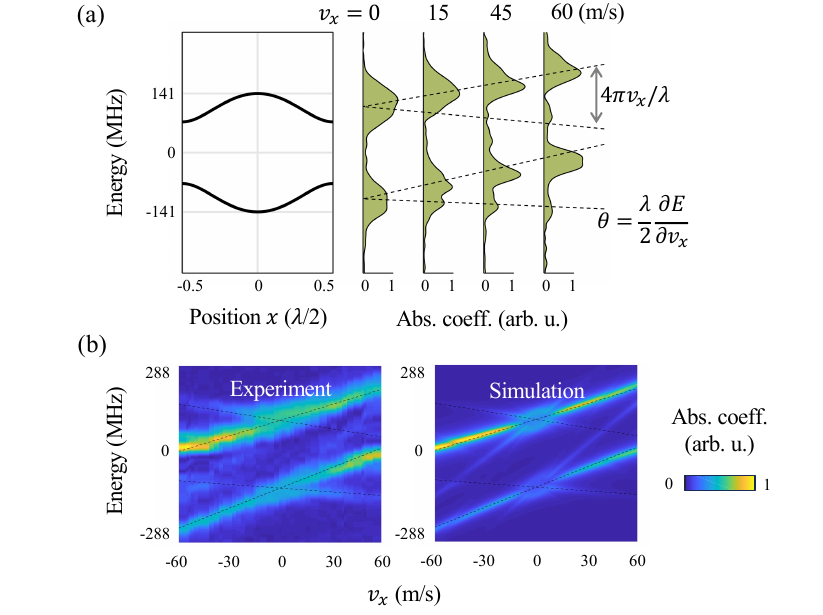}
    \caption{\textbf{Spectroscopic measurement of the Zak phase.} (a) Left panel: the dispersion relation of a gapped two-band SL. Right panel: measured absorption spectra for SLs of atoms with different $v_x$. The energy spectra of WSLs show discrete peaks with equal energy spacing $4\pi v_x/\lambda$. (b) The velocity-resolving absorption spectra of SLs in thermal atoms. For the WSLs from the same band, the slope $\partial E/\partial v_x$ is used to obtain the associated Zak phase. The parameters $t_1=101$ MHz, $t_2=36$ MHz, and $\Delta=-71$ MHz. Dashed lines are theoretically predicted positions of the WSLs.}
        \label{fig2}
\end{figure}

By scanning $\delta$, we perform VST to map out atomic responses as functions of the atomic velocity. For each $\delta$, the absorption spectrum reflects the density of states (DOS) of the SL in the corresponding static electric fields \cite{SM}, which is characterized by the WSLs of the system Hamiltonian $H_x$ (we set $\hbar=1$),
\begin{equation}
H_{1}(x) = \begin{pmatrix}
\Delta/2 & t_1 e^{ikx} + t_2 e^{-ikx}\\
t_1 e^{-ikx} +t_2 e^{ikx} & -\Delta/2 
 \end{pmatrix},
 \label{H_1}
 \end{equation}
in the basis of the states $|a\rangle$ and $|b\rangle$, with $k=2\pi/\lambda$ being the wave vector of the coupling laser.
When $t_1=t_2$ and $\Delta=0$, $H_1$  has two connected bands, which can be considered as a whole band with a real-space Brillouin zone (BZ) length $\lambda$. By setting $v_x=0$, we obtain the DOS of the energy band (see the right panel in Fig.~\ref{fig1}(d)). The two peaks correspond to  the van Hove singularities at the top and bottom of the energy band. For finite velocities $v_x$ in Fig.~\ref{fig1}(d) and (e), the energy band splits into discrete energy levels in WSLs \cite{Wannier1960,Wilkinson1996}. The ladders asymptotically converge to the band center when we reduce  $v_x$ to zero. The WSLs are associated with the fact that the atom periodically passes through the real-space BZ with the dynamic equation
\begin{equation}
x(t) = x_0 + v_x t,
\label{eq:dyn_x} 
\end{equation}
which corresponds to the evolution of crystal electrons in a static electric field, but in momentum space. The Bloch period for this effective single band SL is $T=\lambda/v_x$, which is the time for the excitation returning to its original state, or equivalently, for the atoms moving across a wavelength. Thus, the equal energy spacing of the discrete levels is the corresponding Bloch frequency $\omega = 2\pi/T=2\pi v_x/\lambda$ \cite{Hartmann2004}.


VST also allows for the direct measurement of the geometric Berry phase of the energy bands, i.e., the Zak phase. In Fig.~\ref{fig2}, we intentionally introduce a bandgap by setting $t_1\neq t_2$ and $\Delta\neq 0$. The BZ length of these two-band SLs is $\lambda/2$, leading to a Bloch oscillation period of $T=\lambda/2v_x$ for the moving atoms. This period corresponds to a Bloch oscillation frequency $\omega=4\pi v_x/\lambda$ for both bands, resulting in two sets of WSLs with a ladder spacing twice that of the single-band lattice (see Fig.~\ref{fig1}(d)).
Within the same band, the derivative of the ladder energy $E$ with respect to atomic velocity is proportional to the associated Zak phase (modulo $2\pi$) \cite{Zak1989,Sundaram1999,Lee2015},
\begin{equation}
\theta = \frac{\lambda}{2}\frac{\partial E}{\partial v_x}.
\label{zak}
\end{equation}
This is the central formula for characterizing the topology of energy bands.

\begin{figure}[tp!]
    \centering
    \includegraphics[width =1\linewidth]{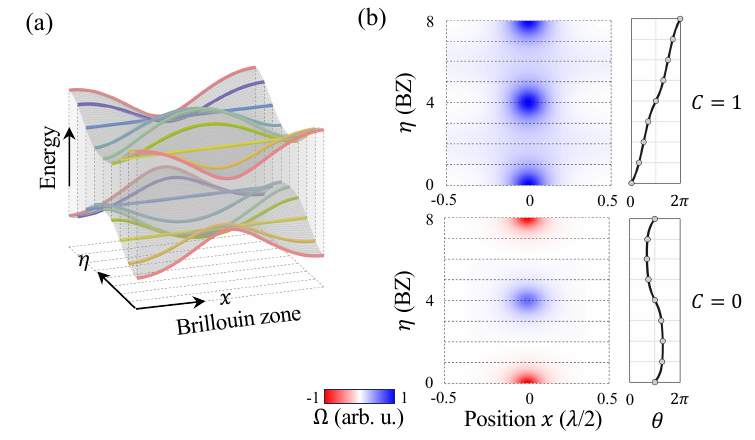}
    \caption{\textbf{Wilson loop approach for measuring the Chern number from the winding of the Zak phase.} (a) The 2D energy  band as a function of $x$ and $\eta$. The band can be divided into a series of 1D bands along $x$, with $\eta$ being a parameter. (b) The Zak phase of $H_1$ (right panels) and the Berry curvature of $H_2$ (left panels). The difference between two adjacent Zak phases equals to the integrated Berry curvature over the enclosed area between the two neighboring dashed lines. Therefore, the winding of the Zak phase with $\eta$ gives the Chern number.
        }
    \label{fig3}
\end{figure}

\begin{figure*}[htp!]
    \centering
    \includegraphics[width = 0.94\linewidth]{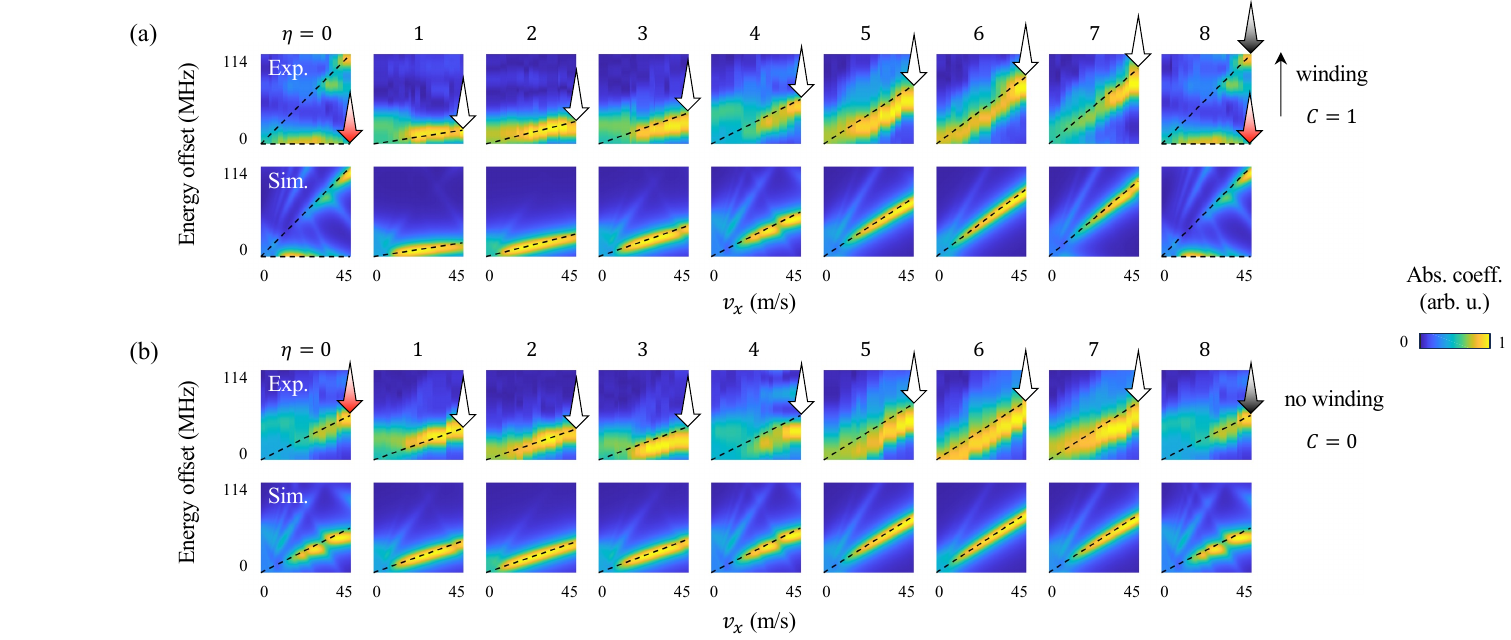}
    \caption{\textbf{Topololgical winding of the Zak phases.} (a) $B=100$ MHz and $u=0$, the Chern number $C=1$. The WSL is shifted by a ladder spacing unit, i.e., has a winding.  (b) $B=12.5$ MHz and $u=87.5$ MHz, the Chern number $C=0$. The WSL returns to its original position without a winding. The upper panels are the VST data for $H_{1}$ with $\eta=0,1,2,..., 8$ from left to right. The lower panels are the corresponding simulation results. The white arrows track the shifts of a major WSL for different $\eta$'s. The red and the black arrows mark the WSL at the two edges ($\eta=0$ and 8) of the 2D BZ.  We fix $A=68$ MHz and $r=2/3$. 
    }
    \label{fig4}
\end{figure*}

VST enables us to measure the Chern number \cite{Klitzing1980,Thouless1982} of a 2D lattice Hamiltonian by counting the topological winding of the Zak phases of a set of 1D SL Hamiltonians through a dimension reduction technique. 
In cold atoms the Chern number is usually obtained by integrating the Berry curvature of the energy bands \cite{Jotzu2014,Aidelsburger2015,Wimmer2017,Chalopin2020}, which, however, requires point-by-point measurement across the entire 2D BZ. 
In the following, we show that VST can be combined with the Wilson loop method \cite{Yu2011,Soluyanov2011,Taherinejad2014,Alexandradinata2014,Benalcazar2017Sci,Benalcazar2017,Xiao2021}, also known as the hybrid Wannier function approach \cite{Marzari2012}, to obtain the Chern numbers of SLs in room-temperature atoms. 

The key strategy of the Wilson loop approach is to divide the 2D BZ into a set of 1D lines (here along the $x$ direction), each line being characterized by a parameter $\eta$ (see Fig.~3(a)). We then measure the Zak phases of the 1D Hamiltonians along these lines.
According to the 
Stokes’ theorem \cite{Abanin2013}, the differences between the Zak phases along the lines with $\eta_1$ and $\eta_2$ is,
\begin{equation}
\theta(\eta_1) - \theta(\eta_2) = \int_{\mathcal{S}}  \Omega d\eta dx,
\end{equation}
where $\Omega$ is the Berry curvature and $\mathcal{S}$ is the BZ area between the two lines. As shown in Fig.~3(b), the Zak phase increases when the enclosed Berry curvature is positive, and vice versa. Therefore, by moving $\eta$ from one edge to the other edge of the BZ, we obtain the total Berry curvature, i.e., the Chern number, of the energy band, from the winding of the Zak phases. By setting $\eta$ as the time $t$, this winding leads to  the well-known Thouless pumping \cite{Benalcazar2017}.

The transport along the 1D lines accumulates Zak phases that are proportional to the displacements of the associated WSLs, according to the modern theory of polarization \cite{King1993,Resta1994} (see Eq.~(\ref{zak})). The winding of the Zak phase with $\eta$ characterizes the adiabatic pumping in the bulk and identifies the Chern number.
As a proof-of-principle demonstration, we introduce a parameter $\eta$ in $H_1$ with $t_1(\eta)= A[1-r\sin(\pi\eta/4)]$, $t_2(\eta)= A[1+r\sin(\pi\eta/4)]$, and $\Delta(\eta) =B\cos(\pi\eta/4)+u$, where $A$, $B$, $r$ and $u$ are parameters that determine the topological property of the 2D Hamiltonian \cite{SM}, 
\begin{equation}
H_{2}(x,\eta) = H_{1}[x, t_1(\eta), t_2(\eta), \Delta(\eta)],
\end{equation}
with $\eta$ being the coordinate of an additional dimension. The effective 2D Hamiltonian $H_2(x,\eta)$ has a period 8 for $\eta$.
The winding of the Zak phase can be obtained by tracking the shift of the WSLs. In Fig.~\ref{fig4}(a), we start from $\eta=0$ at one edge of the 2D BZ and mark a most visible ladder with a red arrow. We then increase $\eta$ and observe that the ladder shifts, as indicated by the white arrows. Finally when $\eta=8$ at the other edge of the BZ, the ladder shifts to the position of the black arrow. Due to the periodicity of the 1D Hamiltonian, the WSLs at the two edges (corresponding to $\eta=0$ and 8) are identical. However, by changing $\eta$ from 0 to 8, the WSLs are shifted by a unit ladder spacing, i.e., the Bloch frequency. The Zak phase undergoes a winding, which corresponds to a Chern number $C=1$ \cite{WangLei2013}. On the contrary, for the parameters in Fig.~\ref{fig4}(b), we observe that the ladders return to their original positions without a winding, demonstrating a Chern number $C=0$.

In summary, we have developed a VST method for accurately measuring the velocity-dependent energy spectra of SLs within thermal atoms. This approach facilitates the extraction of Zak phases associated with the 1D lattice Hamiltonian. By dividing the 2D lattice Hamiltonian into a series of 1D Hamiltonians, we can determine Chern numbers by examining the winding of Zak phases using the Wilson loop method. Our method is adaptable for characterizing various topological materials, including $\mathbb{Z}_2$ topological insulators \cite{Yu2011,Soluyanov2011}, higher-order topological insulators \cite{Benalcazar2017,Benalcazar2017Sci}, and symmetry-protected topological crystalline insulators, which pose challenges for characterization in artificial platforms. Going beyond the dimension reduction technique, we can directly construct two and higher dimensional SLs by introducing multiple coupling fields \cite{Wang2015optica,Zhang2019prl} or Floquet modulation \cite{Xu2022}. The physical quantities can be measured by observing the velocity dependent lattice transport. The Zak phases can also be measured from the transverse pattern of the transmission spectra. This enables the synthesis of diverse exotic topological phases in room-temperature SLs. The integration of VST and the Wilson loop positions SLs as a promising platform for room-temperature quantum simulation. Furthermore, the VST method can also be immediately applied to the atomic frequency comb protocol \cite{Riedmatten2008,Afzelius2010} of quantum memory. The combination of VST with SLs offers a Zak-phase-based method for adjusting the energy spacing between different velocity groups \cite{Saglamyurek2011,Gundogan2012,Zhou2012,Main2021}, thereby providing a control knob for phase-matching time and, consequently, storage durations.

\begin{acknowledgments}
This work was supported by National Natural Science Foundation of China (Grant No. 12325412, 12374335, 11934011, and U21A20437), National Key Research and Development Program of China (Grant No. 2019YFA0308100), Zhejiang Provincial Natural Science Foundation of China (Grant No. LZ24A040001), and the Fundamental Research Funds for the Central Universities.
\end{acknowledgments}

$^{*}$These authors contributed equally to this work.

$^{\dagger}$xuxingqi@zju.edu.cn

$^{\dagger}$hancai@zju.edu.cn

$^{\dagger}$dwwang@zju.edu.cn






\end{document}